\documentclass[runningheads]{svmult}
\usepackage{makeidx}   
\usepackage{graphicx}  
\usepackage{subeqnar}  
\usepackage{multicol}  
\usepackage{physprbb}  
\makeindex             

\newcommand{\greeksym}[1]{{\usefont{U}{psy}{m}{n}#1}}

\newcommand{\udelta}{\mbox{\greeksym{d}}}
\newcommand{\uDelta}{\mbox{\greeksym{D}}}

\begin{document}
\title*{Constraining Variations in the Fine-structure Constant, Quark
  Masses and the Strong Interaction}
\toctitle{Constraining Variations in Fundamental Constants}
%
%
\titlerunning{Constraining Variations in $\alpha$, $m_{\rm q}$ and
$\Lambda_{\rm QCD}$}
%
\author{M.T.~Murphy\inst{1}
\and V.V.~Flambaum\inst{2}
\and J.K.~Webb\inst{2}
\and V.V.~Dzuba\inst{2}
\and J.X.~Prochaska\inst{3}
\and A.M.~Wolfe\inst{4}}
\authorrunning{M.~T.~Murphy, V.~V.~Flambaum et al.}
%
%
\institute{Institute of Astronomy, University of Cambridge, Madingley Road,
  Cambridge CB3 0HA, UK
  \and School of Physics, University of New South Wales, Sydney
  N.S.W. 2052, Australia
  \and UCO-Lick Observatory, University of California, Santa Cruz, Santa
  Cruz, CA 95464, USA
  \and Department of Physics and Centre for Astrophysics and Space
  Sciences, University of California, San Diego, C-0424, La Jolla, CA
  920923, USA}

\label{09}

\maketitle              

\begin{abstract}
We present evidence for variations in the fine-structure constant from
Keck/HIRES spectra of 143 quasar absorption systems over the redshift range
$0.2 < z_{\rm abs} < 4.2$. This includes 15 new systems, mostly at high-$z$
($z_{\rm abs} > 1.8$). Our most robust estimate is a weighted mean
$\uDelta\alpha/\alpha = (-0.57 \pm 0.11) \times 10^{-5}$. We respond to
recent criticisms of the many-multiplet method used to extract these
constraints. The most important potential systematic error at low-$z$ is
the possibility of very different Mg heavy isotope abundances in the
absorption clouds and laboratory: {\it higher} abundances of $^{25,26}$Mg in
the absorbers may explain the low-$z$ results. Approximately equal mixes of
$^{24}$Mg and $^{25,26}$Mg are required. Observations of Galactic stars
generally show {\it lower} $^{25,26}$Mg isotope fractions at the low
metallicities typifying the absorbers. Higher values can be achieved with
an enhanced population of intermediate mass stars at high redshift, a
possibility at odds with observed absorption system element abundances. At
present, all observational evidence is consistent with the varying-$\alpha$
results.

Another promising method to search for variation of fundamental constants
involves comparing different atomic clocks. Here we calculate the
dependence of nuclear magnetic moments on quark masses and obtain limits on
the variation of $\alpha$ and $m_{\rm q}/\!\Lambda_{\rm QCD}$ from recent
atomic clock experiments with hyperfine transitions in H, Rb, Cs, Hg$^+$
and an optical transition in Hg$^+$.
\end{abstract}

\index{Quasar spectra|see{Keck/HIRES~spectra}}
\index{Velocity components|see{Quasar~absorption~lines!velocity~structure}}
\index{VPFIT|see{Quasar~absorption~lines!profile~fitting}}
\index{IMF|see{Initial~mass~function}}
\index{AGB stars|see{Asymptotic~giant~branch~stars}}
\index{DLAs|see{Damped~Lyman-\alpha~systems}}
\index{LLSs|see{Lyman-limit~systems}}
\index{MM method|see{Many-multiplet~method}}
\index{AD method|see{Alkali-doublet~method}}
\index{Abundance studies|see{Element~abundances}}
\index{BBN|see{Big~Bang~nucleosynthesis}}
\index{$\mu$|see{Nuclear~magnetic~moments}}

\section{Introduction}\label{09s:intro}

The last decade has seen the idea of varying fundamental constants receive
unprecedented attention. The historical and modern theoretical motivations
for varying constants, as well as the current experimental constraints, are
reviewed in \cite{UzanJ_03a}, the many articles in \cite{MartinsC_03a} and
this proceedings. Here we set experimental constraints on variations in two
fundamental quantities, the fine-structure constant ($\alpha\equiv
e^2/\hbar c$; Sect.~\ref{09s:quasar}) and the ratio of quark masses to the
quantum chromodynamic (QCD) scale ($m_{\rm q}/\!\Lambda_{\rm QCD}$;
Sect.~\ref{09s:clock}), from optical quasar absorption spectra and
laboratory atomic clocks respectively.

\section{Varying $\alpha$ from Quasar Absorption Lines}\label{09s:quasar}
\index{Variation~of~fundamental~constants!alpha@$\alpha$}

To serve the broad readership of this proceedings, we first outline the
salient features of optical quasar absorption spectroscopy.
Sect.~\ref{09ss:MMmeth} discusses in some detail the recently introduced
\index{Many-multiplet~method}many-multiplet (MM) method of constraining
varying $\alpha$ from optical quasar absorption spectra. All optical data
studied here were obtained at the Keck I 10-m telescope on Mauna
Kea\index{Keck/HIRES~spectra}, Hawaii, with the High Resolution Echelle
Spectrograph \cite{VogtS_94a}. When applied to these data, this is the only
method to date to yield internally robust evidence for a
varying-$\alpha$. We summarize our spectral analysis techniques and present
this evidence in Sect.~\ref{09ss:MMresu}. After responding to some recent
criticisms of the MM method in Sect.~\ref{09ss:MMcrit}, we discuss the most
important potential systematic error for the MM results -- cosmological
isotopic abundance\index{Isotopic~abundances} variations -- in
Sect.~\ref{09ss:MMisot}.

\subsection{Quasar Absorption Lines}\label{09ss:MMqsos}
\index{Quasar~absorption~lines|(}

\begin{figure}[ht!]
\begin{center}
\includegraphics[width=0.85\textwidth]{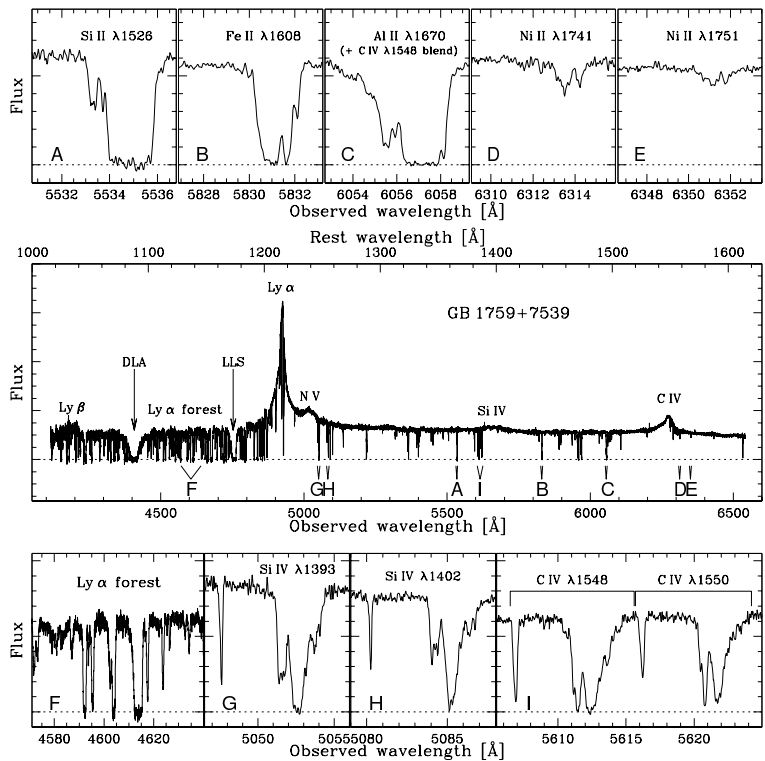}
\end{center}
\vspace{-0.3cm}
\caption[]{Keck/HIRES spectrum of quasar GB 1759$+$7539
  \cite{OutramP_99a}. The full spectrum (middle panel) shows several broad
  emission lines (Ly-$\alpha$, Ly-$\beta$, N{\sc \,v}, C{\sc \,iv}, Si{\sc
  \,iv}) intrinsic to the quasar in the observer's rest-frame (i.e.~vacuum
  heliocentric; bottom) and the quasar rest-frame (top). The dense
  `Lyman-$\alpha$ forest' blue-wards of the Ly-$\alpha$ emission line is
  caused by cosmologically distributed low column-density hydrogen clouds
  along the line of sight to the quasar. The damped Lyman-$\alpha$ system
  (DLA) at $z_{\rm abs}=2.625$ and the Lyman-limit system (LSS) at $z_{\rm
  abs}=2.910$ give rise to heavy-element absorption lines red-wards of the
  Ly-$\alpha$ emission line (away from the confusing Ly-$\alpha$
  forest). Some $z_{\rm abs}=2.625$ transitions are detailed in panels A--E
  \& G--I. Even though the transitions in the top panels have very
  different line-strengths, the velocity structures clearly follow each
  other closely. Detection of many such transitions facilitates
  determination of the velocity structure and allows easy detection of
  random blends. For example, the blue portion of the Al{\sc \,ii}
  $\lambda$1670 profile is blended with C{\sc \,iv} $\lambda$1548 in the
  $z_{\rm abs}=2.910$ LLS}
\index{Keck/HIRES~spectra!examples}
\index{Quasar~absorption~lines!velocity~structure}
\index{Damped~Lyman-\alpha~systems} \index{Lyman-limit~systems}
\index{Lyman-alpha~forest@Lyman-$\alpha$~forest}
\label{09f1}
\end{figure}

The optical spectra of quasars are rich with absorption lines arising from
gas clouds along the line of sight to Earth. For those unfamiliar with the
anatomy of a quasar spectrum, Fig.~\ref{09f1} provides a tutorial. Of
particular interest are those absorption clouds with sufficiently high
hydrogen column density [number of absorbing atoms per area along the line
of sight, $N$(H{\sc \,i})] to make metal absorption lines detectable. These
are classified as Lyman-limit\index{Lyman-limit~systems} and damped
Lyman-$\alpha$ systems\index{Damped~Lyman-\alpha~systems} (LLSs and DLAs):
LLSs have $N$(H{\sc \,i}) $> 2\times 10^{17}{\rm cm}^{-1}$ and DLAs have
$N$(H{\sc \,i}) $> 2\times 10^{20}{\rm cm}^{-1}$. It is the pattern of
metal absorption lines -- the relative separation between the different
transitions -- which carries information about the value of $\alpha$ in the
clouds. The upper and lower panels of Fig.~\ref{09f1} detail some of the
metal lines, those in the upper panel being of particular interest for the
\index{Many-multiplet~method}many-multiplet method in
Sec.~\ref{09ss:MMmeth}. Note the `velocity
structure'\index{Quasar~absorption~lines!velocity~structure} of the
absorption cloud: each transition comprises many `velocity
components'\index{Quasar~absorption~lines!velocity~structure}, each of
which probably corresponds to a separate absorption cloud, all components
probably being associated with a single high redshift galaxy or dark matter
halo.

\index{Quasar~absorption~lines|)}

\subsection{The Many-multiplet (MM) Method}\label{09ss:MMmeth}
\index{Variation~of~fundamental~constants!alpha@$\alpha$|(}
\index{Many-multiplet~method|(}

\index{Alkali-doublet~method|(}
Initial attempts at constraining $\alpha$-variability with quasar
absorption spectra
\cite{SavedoffM_56a,BahcallJ_65a,BahcallJ_67a,WolfeA_76a,CowieL_95a,VarshalovichD_96b}
used the alkali doublet\index{Alkali-doublet~method} (AD) method: for small
variations in $\alpha$, the relative wavelength separation between the
transitions of an AD is proportional to $\alpha$. While the AD method is
simple, it is relatively insensitive to $\alpha$-variations. The $s$ ground
state is most sensitive to changes in $\alpha$ (i.e.~it has the largest
relativistic
corrections\index{Many-multiplet~method!relativistic~corrections}) but is
common to both transitions of the AD. A more sensitive method is to compare
transitions from different multiplets and/or atoms, allowing the ground
states to constrain $\alpha$, i.e.~the many-multiplet (MM) method
introduced in \cite{DzubaV_99a,WebbJ_99a}. We summarize the advantages of
the MM method in \cite{MurphyM_01a}.  \index{Alkali-doublet~method|)}

We first illustrate the MM method with a semi-empirical equation for the
relativistic
correction\index{Many-multiplet~method!relativistic~corrections}, $\uDelta$,
for a transition from the ground state with total angular momentum, $j$:
\index{Many-multiplet~method!relativistic~corrections}
\begin{equation}\label{09eq:delta}
\uDelta \propto (Z\alpha)^2\left[\frac{1}{j+1/2} - C\right]\,,
\end{equation}
where $Z$ is nuclear charge and many-body effects are described by $C \sim
0.6$. To obtain strong constraints on $\alpha$-variability one can (a)
compare transitions of light ($Z \sim 10$) atoms/ions with those of heavy
($Z \sim 30$) ones and/or (b) compare $s$--$p$ and $d$--$p$ transitions of
heavy elements. For the latter, the relativistic
corrections\index{Many-multiplet~method!relativistic~corrections} will be
of opposite sign which further increases sensitivity to $\alpha$-variation
and strengthens the MM method against systematic errors in the quasar
spectra.

In practice, we express the rest-frequency, $\omega_z$, for any transition
observed in the quasar spectra at a redshift $z$, as
\begin{equation}\label{09eq:omega}
\omega_z = \omega_0 + q\left[\left(\frac{\alpha_z}{\alpha}\right)^2 -1
\right] \approx \omega_0 + 2q\uDelta\alpha/\alpha\,,
\end{equation}
where $\alpha_z$ is $\alpha$ in the absorption cloud. For most metal
transitions observed in quasar absorption spectra, the laboratory
wavenumber, $\omega_0$, is measured with low precision compared with that
achievable from the quasar spectra (!) since, in the laboratory, the
transitions fall in the UV. For example, despite a recent order of
magnitude precision gain \cite{GriesmannU_00a}, the C{\sc \,iv}
$\lambda$1548 and 1550 wavenumbers carry formal errors $>0.04{\rm
\,cm}^{-1}$. Compare this with the precision of $\approx\!0.02{\rm
\,cm}^{-1}$ available from absorption lines in a high resolution quasar
spectrum (see Sect.~\ref{09ss:MMcrit}). Dedicated laboratory measurements
\cite{PickeringJ_98a,PickeringJ_00a,GriesmannU_00a} of $\omega_0$ for many
transitions now reach an accuracy of $<0.004{\rm \,cm}^{-1}$ allowing a
precision of $\uDelta\alpha/\alpha \sim 10^{-7}$ to be achieved. Updated
values of $\omega_0$ are given in table 2 of \cite{MurphyM_03a}.

The $q$
coefficient\index{Many-multiplet~method!qcoefficients@$q$~coefficients} of
each transition contains all the relativistic
corrections\index{Many-multiplet~method!relativistic~corrections} and
measures the sensitivity of the transition frequency to changes in
$\alpha$. These have been calculated in
\cite{DzubaV_99a,DzubaV_99b,DzubaV_01a,DzubaV_02a} using many-body
techniques. The accuracy of these calculations is given by how well various
observable quantities (e.g.~spectrum, $g$-factors etc.) of the ion in
question are reproduced. To account for all dominant relativistic effects,
the Dirac-Hartree-Fock
approximation\index{Dirac-Hartree-Fock~approximation} is used as a starting
point. The accuracy is improved using many-body perturbation
theory\index{Many-body~perturbation~theory} and/or the
configuration-interaction method. For most transition combinations used in
the MM method, the accuracy of these calculations is better than 10\%. Note
that in the absence of systematic effects in the quasar spectra, the form
of (\ref{09eq:omega}) ensures one cannot infer a non-zero
$\uDelta\alpha/\alpha$ due to errors in the $q$
coefficients\index{Many-multiplet~method!qcoefficients@$q$~coefficients}. The
$q$ coefficients used in this paper are compiled in table 2 of
\cite{MurphyM_03a}.

\begin{figure}[ht!]
\begin{center}
\hbox{
  \includegraphics[width=0.49\textwidth,height=6.0cm]{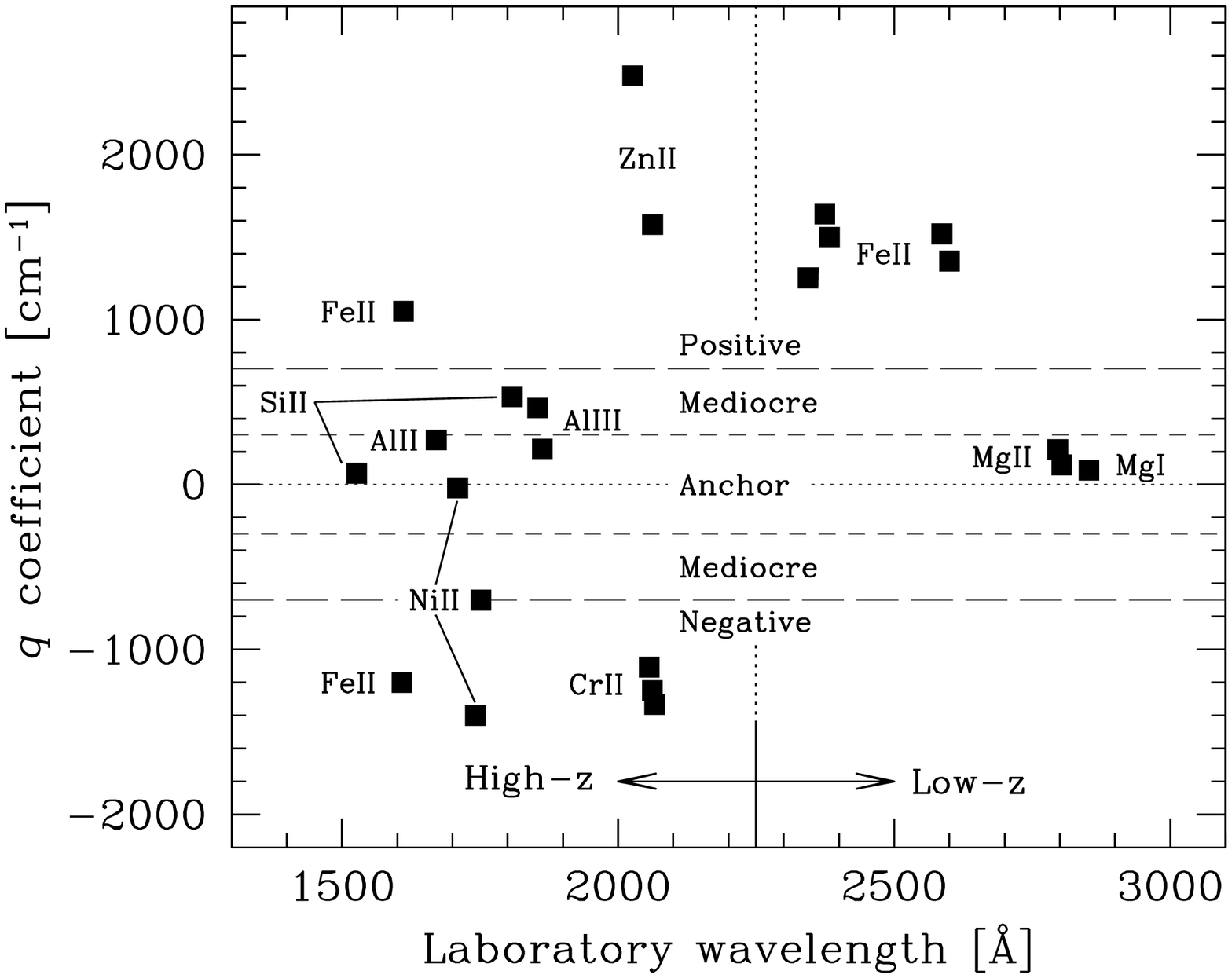}
  \includegraphics[width=0.49\textwidth,height=6.0cm]{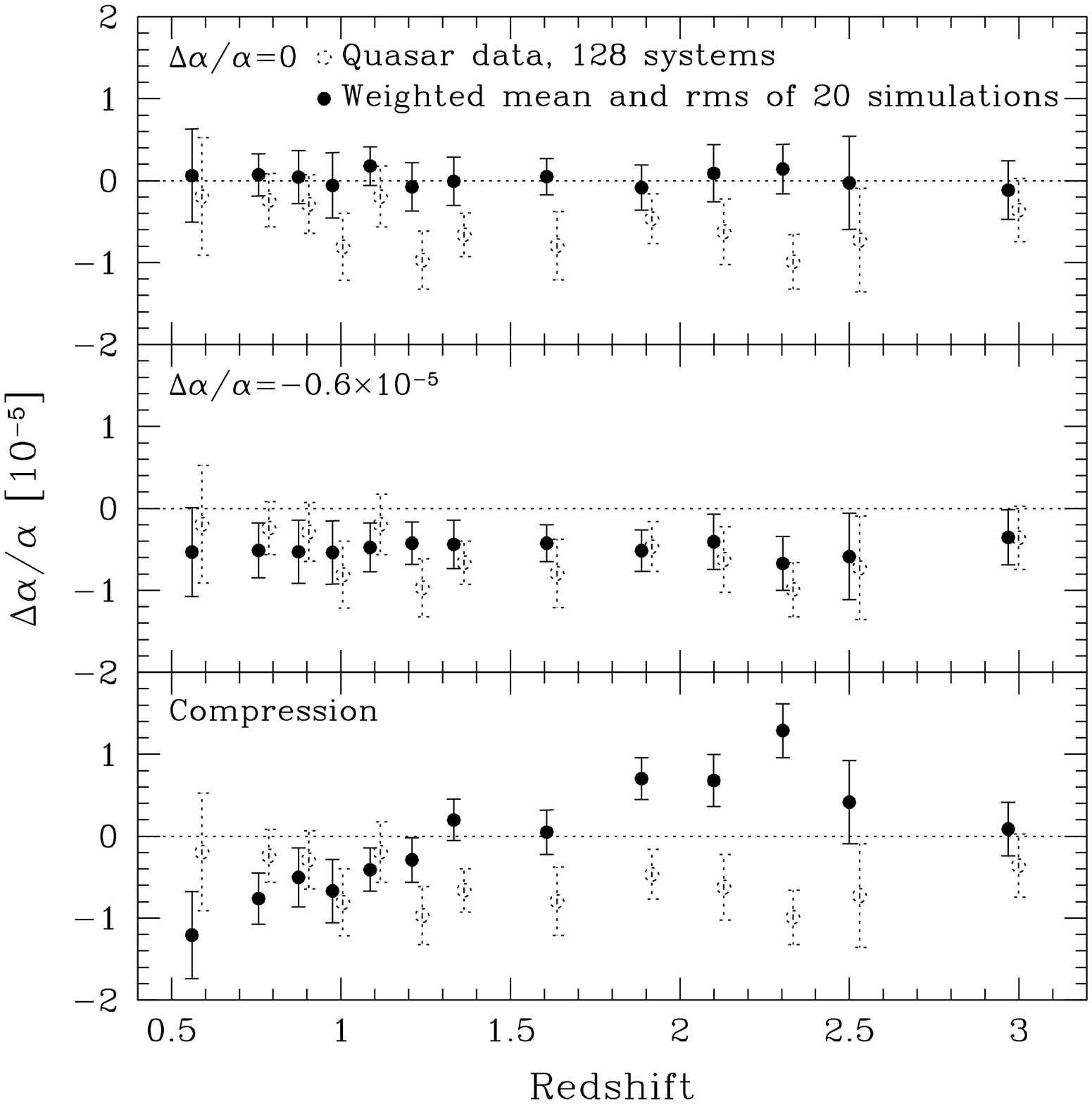}
}
\end{center}
\vspace{-0.6cm}
\caption[]{({\bf Left}) Distribution of $q$ coefficients for transitions
  used in the MM method. For low-$z$ Mg/Fe{\sc \,ii} systems, a compression
  of the spectrum can mimic $\uDelta\alpha/\alpha <0$. However, the complex
  arrangement at high-$z$ indicates resistance to such systematics. We
  define several `$q$-types' by the horizontal bands and labels
  shown. ({\bf Right}) Binned measurements of $\uDelta\alpha/\alpha$ from
  20 simulations of 128 absorption systems ({\it solid}) and the same, real
  quasar absorption systems ({\it dotted}). For the top and middle panels
  we input the indicated values of $\uDelta\alpha/\alpha$. The values and
  errors are recovered reliably. A wavelength compression is introduced for
  the bottom panel to reproduce the low-$z$ quasar results. At high-$z$,
  the variety of $q$ coefficients causes the expected large scatter but the
  average effect on $\uDelta\alpha/\alpha$ is opposite to that in the
  low-$z$ systems. Simple distortions of the quasar spectra cannot explain
  the results}
\index{Many-multiplet~method!qcoefficients@$q$~coefficients}
\index{Many-multiplet~method!qtype@$q$-type}
\index{Keck/HIRES~spectra!simulations}
\label{09f2}
\end{figure}

Figure \ref{09f2} shows the distribution of $q$
coefficients\index{Many-multiplet~method!qcoefficients@$q$~coefficients} in
(rest) wavelength space. Our sample conveniently divides into low- and
high-$z$ subsamples with very different properties (throughout this work we
define $z<1.8$ as low-$z$ and $z>1.8$ as high-$z$). Note the simple
arrangement for the low-$z$ Mg/Fe{\sc \,ii} systems: the Mg transitions are
used as anchors against which the large, positive shifts in the Fe{\sc
\,ii} transitions can be measured. Compare this with the complex
arrangement for the high-$z$ systems: low-order distortions in the
wavelength scale of the quasar spectra will have a varied and complex
effect on $\uDelta\alpha/\alpha$ depending on which transitions are fitted
in a given absorption system. This complexity at high-$z$ may yield more
robust estimates of $\uDelta\alpha/\alpha$. The right panel quantifies this
using simulations\index{Keck/HIRES~spectra!simulations} of the original 128
absorption system sample (see next section) which have been artificially
compressed (see \cite{MurphyM_03a} for details). Even though the systems at
high-$z$ each respond differently to the compression of the wavelength
scale, the binned plot reveals the average response is opposite to that in
the low-$z$ Mg/Fe{\sc \,ii} systems. This is an important strength of the
MM method: the low- and high-$z$ samples respond differently to simple
systematic errors due to their different arrangement of $q$
coefficients\index{Many-multiplet~method!qcoefficients@$q$~coefficients} in
wavelength space.

\index{Variation~of~fundamental~constants!alpha@$\alpha$|)}
\index{Many-multiplet~method|)}

\subsection{Spectral Analysis and Updated Results}\label{09ss:MMresu}

\subsection*{Sample Definition}\label{09sss:MMsamp}
\index{Keck/HIRES~spectra!sample~definition|(}

\begin{figure}[ht!]
\begin{center}
\vbox{
  \includegraphics[height=0.70\textwidth,angle=270]{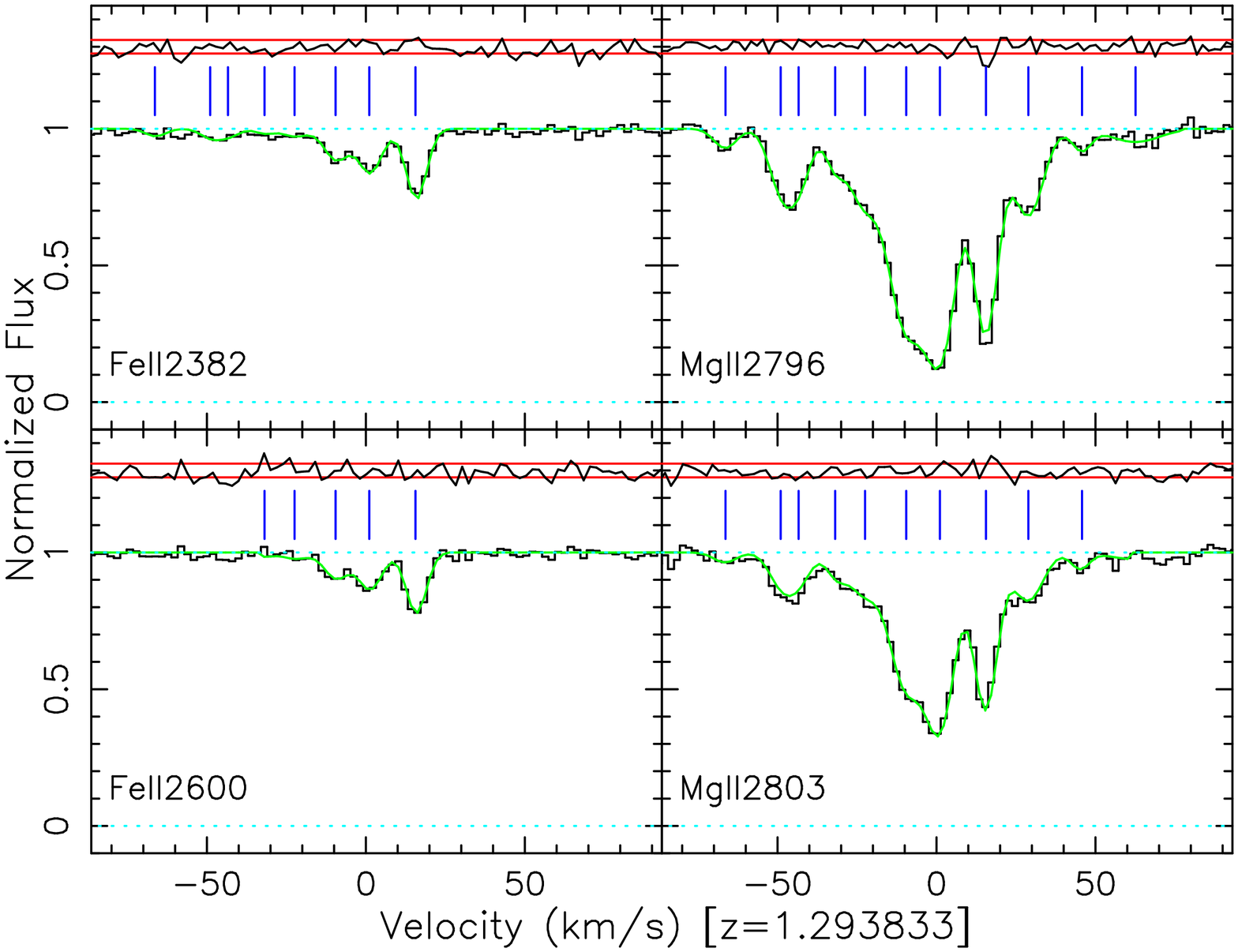}\vspace{0.3cm}
  \includegraphics[height=0.70\textwidth,angle=270]{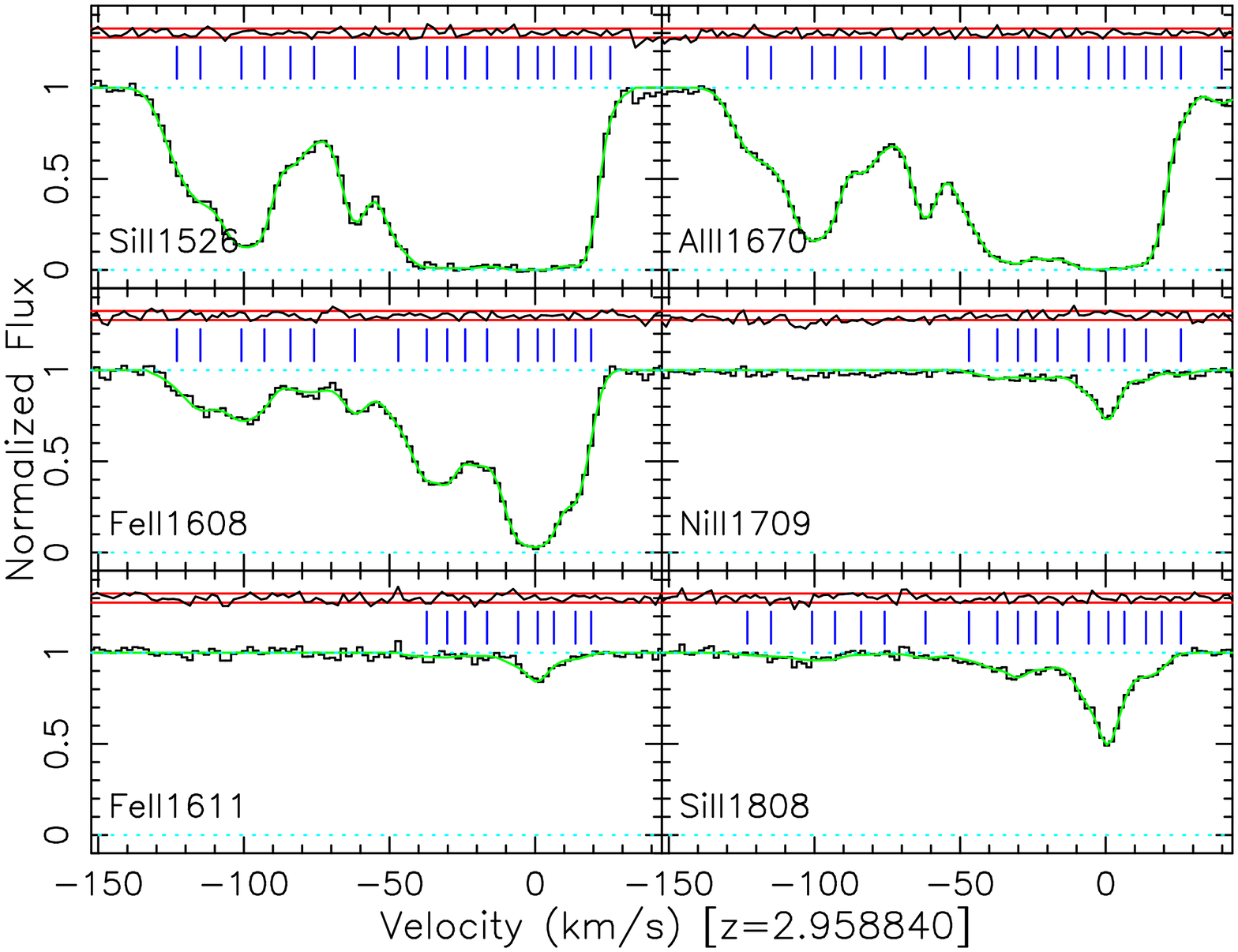}
}
\end{center}
\vspace{-0.3cm}
\caption[]{Selected systems and transitions registered on a common
  (arbitrary) velocity scale. Data ({\it histogram}) are normalized by a
  fitted continuum. Our Voigt profile fit ({\it solid curve}) and
  residuals, normalized to the $1\sigma$ errors ({\it horizontal solid
  lines}), are also shown. Tick-marks indicate individual velocity
  components. ({\bf Top}) $z_{\rm abs}=1.2938$ Mg/Fe{\sc \,ii} system
  towards Q0636$+$6801. ({\bf Bottom}) $z_{\rm abs}=2.9587$ DLA towards
  Q1011$+$4315. Note that only optically thin components constrain
  $\uDelta\alpha/\alpha$ strongly.}
\index{Quasar~absorption~lines!velocity~structure}
\index{Quasar~absorption~lines!profile~fitting}
\index{Keck/HIRES~spectra!examples}
\index{Damped~Lyman-\alpha~systems}
\label{09f3}
\end{figure}

Our data set comprises three samples of Keck/HIRES spectra, each observed
independently by different groups. The first sample \cite{ChurchillC_00a}
contains 27 low-$z$ Mg/Fe{\sc \,ii} systems. The second sample
\cite{ProchaskaJ_99b} contains transitions from a wide variety of ionic
species (though mostly singly ionized; Al{\sc \,ii}, Al{\sc \,iii}, Si{\sc
\,ii}, Cr{\sc \,ii}, Fe{\sc \,ii}, Ni{\sc \,ii} and Zn{\sc \,ii}) in 19
high-$z$ DLAs\index{Damped~Lyman-\alpha~systems} and 3 low-$z$ Mg/Fe{\sc
\,ii} systems. An additional high-$z$ DLA is from \cite{OutramP_99a}. The
third sample was graciously provided by W.~L.~W.~Sargent and collaborators
and comprises 78 absorption systems over a wide redshift range. Together,
these samples comprise 128 absorption systems and form the total sample
presented in \cite{MurphyM_03a}. To illustrate many points in the following
sections, we provide example spectra of a low-$z$ Mg/Fe{\sc \,ii} system
and a high-$z$ DLA\index{Damped~Lyman-\alpha~systems} in Fig.~\ref{09f3}.

In this work we update the second sample with 15 additional systems
observed and reduced by two of the authors (JXP \& AMW) and collaborators
\cite{ProchaskaJ_01a}, containing a mix of low- and high-$z$ systems.

From the fully reduced spectra we select all systems which contain at least
2 transitions of different
$q$-type\index{Many-multiplet~method!qtype@$q$-type} (defined in
Fig.~\ref{09f2}), thereby potentially providing a tight constraint on
$\uDelta\alpha/\alpha$. Only in cases where all selected transitions have
very low signal-to-noise ratio (SN) do we not attempt a
fit\index{Quasar~absorption~lines!profile~fitting}. Only in very high SN
cases have we selected systems where only transitions of the same
$q$-type\index{Many-multiplet~method!qtype@$q$-type} are detected. Apart
from the obvious issues of line-strength and possible random blends
(Sect.~\ref{09ss:MMcrit}), many well known instrumental limitations prevent
us from detecting all MM transitions in every system. For example, the
throughput of the telescope/spectrograph and detector sensitivity drops
sharply below 4000\,\AA\ and above 6000\,\AA. Typically, the spectrum is
not recorded below 3500\,\AA\ and above 7000\,\AA. Also, gaps in the
wavelength coverage appear, particularly towards the red, since the
spectrograph is an echelle--cross-disperser combination. Echelle
orders\index{Echelle~order} cover $\sim 60{\rm \,\AA}$ and inter-order gaps
can be up to $\sim 20{\rm \,\AA}$.

\index{Keck/HIRES~spectra!sample~definition|)}

\subsection*{Profile Fitting}\label{09sss:MMprof}
\index{Quasar~absorption~lines!profile~fitting|(}

For each system, the available transitions are fitted with multiple
velocity component\index{Quasar~absorption~lines!velocity~structure} Voigt
profiles. Each velocity
component\index{Quasar~absorption~lines!velocity~structure} is described by
three parameters: the absorption redshift $z_{\rm abs}$, the Doppler
broadening or $b$ parameter and the column density $N$. We reduce the
number of free parameters by assuming either a completely turbulent or
completely thermal broadening mechanism: corresponding components in all
transitions have equal $b$ or their $b$s are related by the inverse
square-root of the ion masses. To apply the MM method one must assume that
corresponding velocity
components\index{Quasar~absorption~lines!velocity~structure} in all fitted
ions have the same redshift. This further reduces the number of free
parameters. We discuss this assumption in Sect.~\ref{09ss:MMcrit}. To each
{\it system} a single extra parameter is added, $\uDelta\alpha/\alpha$. This
allows all velocity
components\index{Quasar~absorption~lines!velocity~structure} to shift in
concert according to their $q$
coefficients\index{Many-multiplet~method!qcoefficients@$q$~coefficients}.

All free parameters are determined simultaneously using {\sc vpfit}, a
non-linear least-squares $\chi^2$ reduction algorithm written specifically
for analysis of quasar absorption spectra. The 1\,$\sigma$ parameter
uncertainties are determined in the usual way from the diagonal terms of
the final parameter covariance matrix. The assumption that off-diagonal
terms are small (that parameters are not closely correlated) is a good one
for $\uDelta\alpha/\alpha$: redshift and $\uDelta\alpha/\alpha$ are not
correlated (Fig.~\ref{09f2}). Monte Carlo
simulations\index{Keck/HIRES~spectra!simulations} with 10000 realisations
confirm the reliability of the parameter and error estimates
\cite{MurphyM_02b}. It is important to realise that this numerical method
ensures that constraints on $\uDelta\alpha/\alpha$ are derived in a natural
way from {\it optically thin} lines and not from saturated ones. The
derivatives of $\chi^2$ with respect to the saturated component redshifts
are very small compared to the optically thin case and so only the
optically thin lines strongly constrain $\uDelta\alpha/\alpha$. If the two
broadening mechanisms mentioned above result in significantly different
$\uDelta\alpha/\alpha$, the system is rejected.  Otherwise, the broadening
mechanism giving the lowest $\chi^2$ fit is selected. We also require that
$\chi^2$ per degree of freedom, $\chi^2_\nu$, is $\approx 1$.

\index{Quasar~absorption~lines!profile~fitting|)}

\subsection*{Results}\label{09sss:MMresu}
\index{Variation~of~fundamental~constants!alpha@$\alpha$|(}
\index{Many-multiplet~method!results|(}

\begin{figure}[ht!]
\begin{center}
\includegraphics[width=0.65\textwidth]{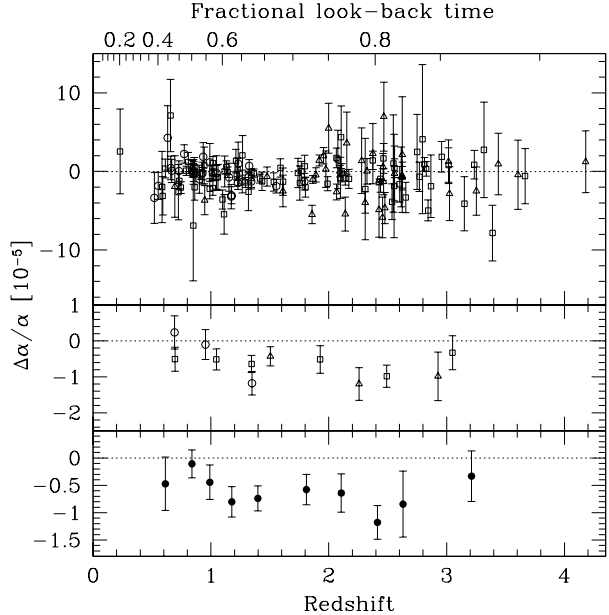}
\end{center}
\vspace{-0.3cm}
\caption[]{$\uDelta\alpha/\alpha$ and 1\,$\sigma$ errors from the
  many-multiplet method for three Keck/HIRES samples: Churchill ({\it
  hollow circles}), Prochaska \& Wolfe ({\it triangles}), Sargent ({\it
  squares}). Upper panel: unbinned individual values. Middle panel: binned
  results for each sample. Lower panel: binned over the whole sample. To
  calculate the fractional look-back time we use $H_0=70{\rm
  \,km\,s}^{-1}{\rm Mpc}^{-1}$, $\Omega_{\rm m}=0.3$, $\Omega_\Lambda=0.7$,
  implying an age of $13.47{\rm \,Gyr}$.}
\label{09f4}
\end{figure}

\begin{table}
\caption{Statistics for different samples. $\chi^2_\nu$ is $\chi^2/(N_{\rm
    sys}-1)$ about the weighted mean}
\begin{center}
\renewcommand{\arraystretch}{1.0}
\setlength\tabcolsep{15pt}
\begin{tabular}{@{}lcccr}
\hline\noalign{\smallskip}
Sample                   & $N_{\rm sys}$ & $\left<z_{\rm abs}\right>$ & $\uDelta\alpha/\alpha$ ($10^{-5}$) & $\chi^2_\nu$ \\
\noalign{\smallskip}
\hline
\noalign{\smallskip}
Churchill                & 27            & 1.00                        & $-0.531 \pm 0.223$                 & 1.109        \\
Prochaska \& Wolfe       & 38            & 2.27                        & $-0.664 \pm 0.219$                 & 2.024        \\
Sargent                  & 78            & 1.76                        & $-0.620 \pm 0.129$                 & 1.182        \\
Low-$z$ ($z<1.8$)        & 77            & 1.07                        & $-0.537 \pm 0.124$                 & 1.064        \\
High-$z$ ($z>1.8$)       & 66            & 2.55                        & $-0.744 \pm 0.167$                 & 1.739        \\
Raw total                &143            & 1.75                        & $-0.611 \pm 0.100$                 & 1.373        \\
{\bf Fiducial}$^{\rm a}$ &{\bf 143}      & {\bf 1.75}                  & ${\bf -\!0.573 \pm 0.113}$         & {\bf  1.023} \\
\noalign{\smallskip}
\hline
\end{tabular}
\end{center}
$^{\rm a}$ Low-$z$ sample $+$ low-contrast sample $+$ high-contrast sample
with increased errors.
\label{09tab:MMresu}
\end{table}

The distribution of $\uDelta\alpha/\alpha$ with redshift and look-back time
as a fraction of the age of the Universe is shown in Fig.~\ref{09f4}. We
also provide basic statistics for the different samples and the total, raw
sample as a whole in Table \ref{09tab:MMresu}. Note that all three samples
give consistent, significantly smaller values of $\alpha$ in the absorption
clouds compared to the laboratory. Breaking the sample down into low- and
high-$z$ subsamples also yields consistent results despite the very
different $q$
coefficient\index{Many-multiplet~method!qcoefficients@$q$~coefficients}
combinations used (Fig.~\ref{09f2}, left) and overall reaction to simple
systematic errors (Fig.~\ref{09f2}, right). We have conducted numerous
internal consistency checks on these results, including direct tests of the
wavelength calibration of the quasar spectra and the effect of removing
individual transitions or entire ionic species from our fits. These are
described in detail in \cite{MurphyM_01a,MurphyM_03a}.

\index{Many-multiplet~method!results|)}

\subsection*{Extra Scatter at High $z$}\label{09sss:MMscat}

Note that the scatter in the total low-$z$ sample is consistent with that
expected from the size of the error bars (i.e.~$\chi^2_\nu \approx 1$).
However, at high-$z$, Fig.~\ref{09f4} shows significant extra scatter. This
is reflected in the high $\chi^2_\nu$ values in Table
\ref{09tab:MMresu}. The weighted mean therefore exaggerates the true
significance of $\uDelta\alpha/\alpha$ at high-$z$.

We have identified the major source of this extra scatter at high
$z$. Consider fitting\index{Quasar~absorption~lines!profile~fitting} two
transitions with very different line-strengths (e.g.~Al{\sc \,ii}
$\lambda$1670 and Ni{\sc \,ii} $\lambda$1709 in Fig.~\ref{09f3}). Weak
components near the high optical depth edges of the strong transition's
profile are not necessary to obtain a good fit to the data. Even though the
{\sc vpfit}\index{Quasar~absorption~lines!profile~fitting} $\chi^2$
minimization ensures that constraints on $\uDelta\alpha/\alpha$ derive
primarily from the optically thin velocity
components\index{Quasar~absorption~lines!velocity~structure}, these weak
components missing from the fit will cause small line shifts. The resulting
shift in $\uDelta\alpha/\alpha$ is random from component to component and
from system to system: the effect of missing components will be to increase
the random scatter in the individual $\uDelta\alpha/\alpha$ values. This
effect will be far larger in the high-$z$ sample since only there do we fit
transitions of such different line-strengths.

\begin{figure}[ht!]
\begin{center}
\includegraphics[width=0.65\textwidth]{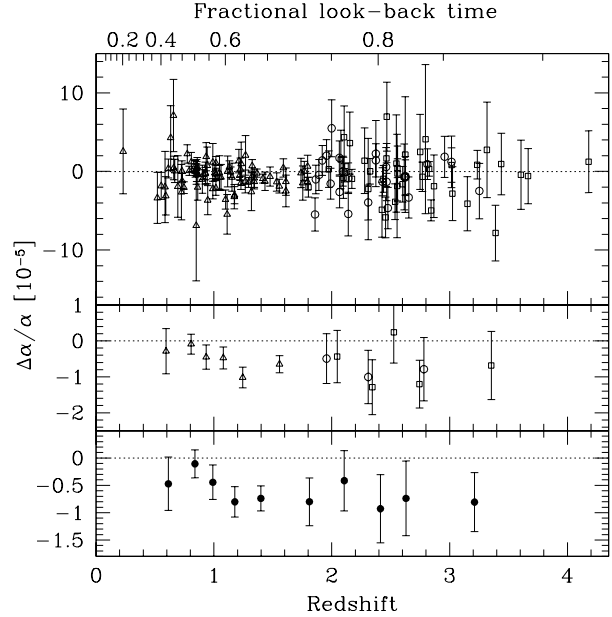}
\end{center}
\vspace{-0.3cm}
\caption[]{The fiducial sample. Low-$z$ sample ({\it triangles}),
  low-contrast sample ({\it squares}) and high-contrast sample with
  increased error bars ({\it circles}). The weighted mean,
  $\uDelta\alpha/\alpha = (-0.57 \pm 0.11)\times 10^{-5}$, is our most
  robust estimate for the HIRES spectra}
\label{09f5}
\end{figure}

We form a `high-contrast' sample from the high-$z$ (i.e.~$z_{\rm abs} >
1.8$) sample by selecting systems in which we
fit\index{Quasar~absorption~lines!profile~fitting} both strong and weak
lines, i.e.~any of the Al{\sc \,ii}, Si{\sc \,ii} or Fe{\sc \,ii}
transitions {\it and} any of the Cr{\sc \,ii}, Ni{\sc \,ii} or Zn{\sc \,ii}
ones. To obtain a more robust estimate of the significance of
$\uDelta\alpha/\alpha$ in this high-contrast sample, we have increased the
individual 1\,$\sigma$ errors until $\chi^2_\nu = 1$ about the weighted
mean. We achieve this by adding $1.75 \times 10^{-5}$ in quadrature to the
error bars of the 27 relevant systems. Other procedures for estimating the
significance at high-$z$ are discussed in
\cite{MurphyM_03a}. Fig.~\ref{09f5} identifies the high-contrast sample and
presents the binned results with increased error bars. Table
\ref{09tab:MMresu} presents the relevant statistics. The above procedure
results in our most robust estimate from the 143 absorption systems over
the redshift range $0.2 < z_{\rm abs} < 4.2$: $\uDelta\alpha/\alpha = (-0.57
\pm 0.11)\times 10^{-5}$.

\index{Variation~of~fundamental~constants!alpha@$\alpha$|)}

\subsection{Recent Criticisms of the MM Method}\label{09ss:MMcrit}
\index{Many-multiplet~method!criticisms|(}

Bekenstein \cite{BekensteinJ_03a} has pointed out that the form of the
Dirac Hamiltonian for the hydrogen atom changes in a theory where $\alpha$
is considered a dynamical field. He warns that the resulting shifts in
energy levels could be important for the MM method. Reference
\cite{AngstmannE_03a} have extended Bekenstein's model to many-electron
atoms. They find that the energy shift within the Bekenstein model is
proportional to $\uDelta\alpha/\alpha$ defined in (\ref{09eq:omega}), up to
corrections of order 1\%. Thus, modifications to the energy shifts
discussed by \cite{BekensteinJ_03a} for the hydrogen atom are not important
for the MM method applied to many-electron metal ions.

Bahcall et al.~\cite{BahcallJ_03a} criticized the MM method on many points,
summarized in their table 3. Though none of these are real candidate
explanations of our results, we address each criticism below to avoid
future confusion:
\begin{itemize}
\item {\bf Theory:} Since one must calculate the $q$
coefficients\index{Many-multiplet~method!qcoefficients@$q$~coefficients}
using sophisticated many-body techniques, \cite{BahcallJ_03a} argue that
the MM method is less reliable than, say, the AD method. The likely sources
of error are discussed in detail by \cite{DzubaV_02a} and we give a flavour
of them in Sect.~\ref{09ss:MMmeth}. The $q$
coefficients\index{Many-multiplet~method!qcoefficients@$q$~coefficients}
are known to high enough accuracy given our sample precision (discussed
below). We again stress that if $\uDelta\alpha/\alpha$ is really zero, one
cannot manufacture a non-zero value through errors in the $q$
coefficients\index{Many-multiplet~method!qcoefficients@$q$~coefficients}
(\ref{09eq:omega}) if systematic errors in the quasar spectra are not
important.
\item {\bf Absolute or relative wavelengths?:} Reference
\cite{BahcallJ_03a} argued that the MM method requires the measurement of
absolute wavelengths in the quasar spectra\footnote{This suggestion is
present only in preprint versions 1 \& 2 of \cite{BahcallJ_03a}
(astro-ph/0301507). It is removed from later versions. Despite this, we
address this point here to avoid further confusion in the
literature.}. This is incorrect. Since we simultaneously determine the
redshifts of the absorption components and $\uDelta\alpha/\alpha$ (these are
not degenerate parameters; Fig.~\ref{09f2}) for each system, any velocity
shift may be applied to the spectra and $\uDelta\alpha/\alpha$ will be
unaffected. A velocity-space shift is a systematic error for the absorption
redshifts, not for $\uDelta\alpha/\alpha$.
\item {\bf Sample precision:} It is trivial to estimate the expected
precision available from our observational sample. We explained this
calculation in \cite{MurphyM_01a} and repeat it here. The
Keck/HIRES\index{Keck/HIRES~spectra!simulations} pixels cover $\sim\!3{\rm
\,km\,s}^{-1}$ and so one reasonably expects to centroid barely resolved
features (with ${\rm SN}\sim$~30--50) to $0.3{\rm \,km\,s}^{-1}$. One
expects $\sim$\,4 features per absorption system to be well-centroided in
this way, providing a velocity precision of $\uDelta v\approx 0.15{\rm
\,km\,s}^{-1}$ or $\uDelta\omega\approx 0.02{\rm \,cm}^{-1}$ for an
$\omega_0\approx 40000{\rm \,cm}^{-1}$ transition. The typical difference
in $q$ coefficient\index{Many-multiplet~method!qcoefficients@$q$~coefficients} between the Mg{\sc \,i}/{\sc ii} and Fe{\sc \,ii} lines
is $\sim\!1000{\rm \,cm}^{-1}$ and so, for a single Mg/Fe{\sc \,ii}
absorption system, (\ref{09eq:omega}) implies a precision of
$\left|\uDelta\alpha/\alpha\right| \sim 1\times 10^{-5}$. With $\sim$\,50
such systems, one expects a precision $\left|\uDelta\alpha/\alpha\right|
\sim 0.14\times 10^{-5}$ (cf.~Table \ref{09tab:MMresu}). As shown in
Fig.~\ref{09f2} (and figs.~A.2 \& A.3 in \cite{MurphyM_02b}),
simulations\index{Keck/HIRES~spectra!simulations} also provide a simple
``reality-check'' on the sample precision.
\item {\bf Line misidentification and blending:} Reference
\cite{BahcallJ_03a} argue we may have misidentified many absorption
features. In high resolution ($R\sim 50000$) spectra we largely resolve the
velocity structure\index{Quasar~absorption~lines!velocity~structure} of
absorption systems. Misidentifying transitions is highly improbable since,
even by eye, the profiles of different species follow each other to within
$\left|\uDelta v\right| < 1{\rm \,km\,s}^{-1}$.  Confirming this, we obtain
good fits\index{Quasar~absorption~lines!profile~fitting} to the absorption
profiles with the number of free parameters restricted by physical
considerations (Sect.~\ref{09sss:MMprof}). Detecting blends from absorption
at other redshifts is also greatly facilitated by high resolution. See
\cite{MurphyM_01a,MurphyM_03a} for thorough discussions of blending and
Fig.~\ref{09f1} for an example. Even if we misidentified a small number of
transitions in our sample and they miraculously mimicked the velocity
structure\index{Quasar~absorption~lines!velocity~structure} of other
detected transitions (thereby allowing a good fit), this would have a
random, not systematic, effect on $\uDelta\alpha/\alpha$. Indeed, compared
with the AD method, the MM method is distinctly robust against
misidentifications and blends: many transitions constrain the velocity
structure\index{Quasar~absorption~lines!velocity~structure} so identifying
blends and misidentifications is all the more trivial. Also, any blends
that are not identified have a smaller effect on $\uDelta\alpha/\alpha$
since many other transitions contribute to the constraints.
\item {\bf Velocity
structure:}\index{Quasar~absorption~lines!velocity~structure}
\index{Quasar~absorption~lines!profile~fitting} The MM method assumes that
corresponding velocity components in different ions have the same
redshift. Most transitions used are from ionic species with very similar
ionization potentials and so absorption from these species arise should
arise co-spatially. Consider a Mg{\sc \,ii} velocity
component\index{Quasar~absorption~lines!velocity~structure} {\it
blue}shifted with respect to the {\it corresponding} Fe{\sc \,ii} component
by some kinematic effect in the absorption cloud. Clearly, this mimics
$\uDelta\alpha/\alpha < 0$ for that component. However, kinematic effects
would equally well {\it red}shift the Mg{\sc \,ii} components. Thus, the
effect on $\uDelta\alpha/\alpha$ is random from component to component and
absorption system to system. This argument is misunderstood in
\cite{BahcallJ_03a}: they feel it is unlikely that such random effects
``average out to an accuracy of $0.2{\rm \,km\,s}^{-1}$ over a velocity
range of more than $10^2{\rm \,km\,s}^{-1}$'', the latter quantity
referring to the total velocity extent of a typical absorption {\it
system}. Inspecting Fig.~\ref{09f4}, it is clear that any extra scatter in
$\uDelta\alpha/\alpha$ from kinematic effects derives only from the gas
properties on velocity scales less than typical $b$ parameters, i.e.~$<
5{\rm \,km\,s}^{-1}$. That we obtain excellent agreement between the
velocity component\index{Quasar~absorption~lines!velocity~structure}
redshifts in different species to a precision $\left|\uDelta v\right| \sim
0.3{\rm \,km\,s}^{-1}$ illustrates this. If kinematic effects were
important, they would be most prominent in the low-$z$ values of
$\uDelta\alpha/\alpha$ from the Mg/Fe{\sc \,ii} systems, appearing as an
extra scatter beyond that expected from the 1\,$\sigma$ errors. This is not
observed. We discuss kinematic effects in more detail in \cite{WebbJ_03b}
and \cite{MurphyM_03a}.
\end{itemize}

\index{Many-multiplet~method!criticisms|)}

\subsection{Isotopic Abundance Variations}\label{09ss:MMisot}
\index{Isotopic~abundances|(}

\begin{figure}[ht!]
\begin{center}
\hbox{
  \hspace{-0.1cm}\includegraphics[width=0.327\textwidth]{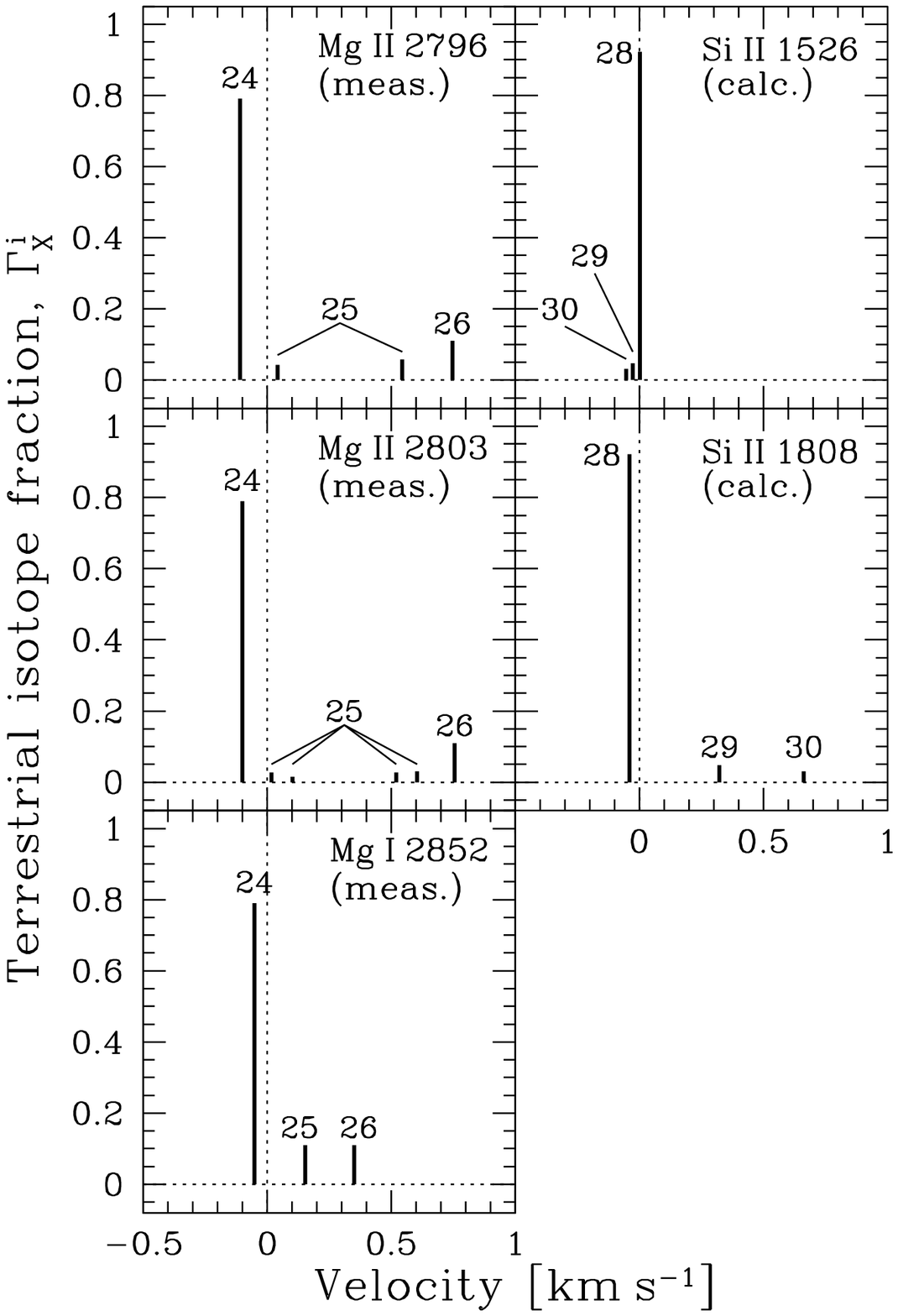}
  \includegraphics[width=0.66\textwidth]{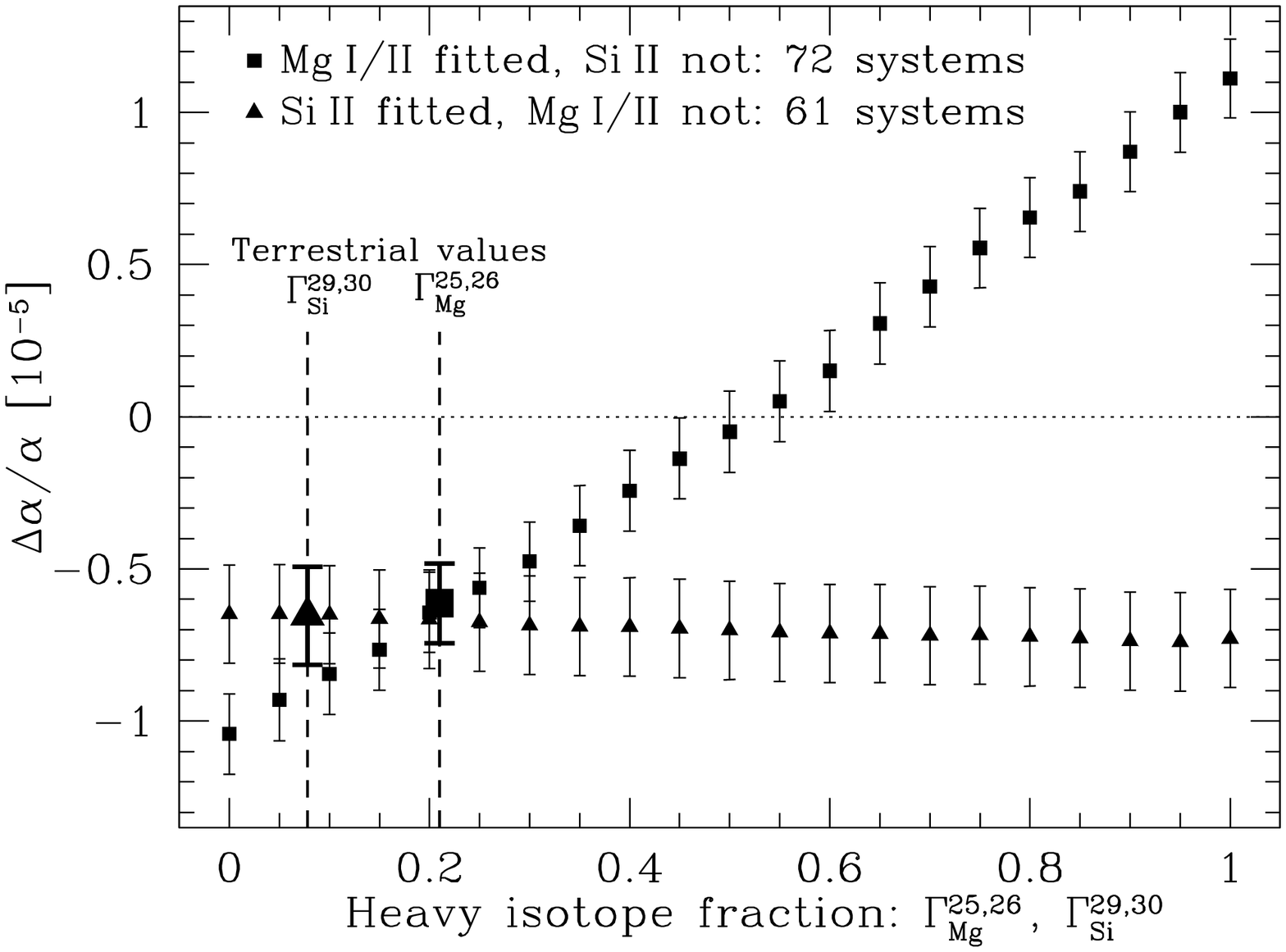}
}
\end{center}
\vspace{-0.6cm}
\caption[]{({\bf Left}) Isotopic structures for relevant Mg/Si transitions
  from measurements \cite{HallstadiusL_79a,DrullingerR_80a,PickeringJ_98a}
  or calculations \cite{BerengutJ_03a}. Zero velocity corresponds to the
  structure's centre of gravity. ({\bf Right}) Sensitivity of
  $\uDelta\alpha/\alpha$ to isotopic abundance variations. We alter the
  heavy isotope abundances proportionately: the heavy element fraction for
  Mg is $\Gamma^{25,26}_{\rm Mg} = {\rm const.}\times (0.10 + 0.11)$ where
  the numbers in parentheses are the terrestrial isotopic fractions of
  $^{25}$Mg and $^{26}$Mg. Much higher relative $^{25,26}$Mg abundances in
  the absorbers can explain the low-$z$ results but the high-$z$ results
  (containing the Si{\sc \,ii} systems shown) are insensitive to
  $^{29,30}$Si abundances} \index{Isotopic~structure}
\label{09f6}
\end{figure}

As we discussed in \cite{MurphyM_01b,WebbJ_01a} and emphasised in
\cite{MurphyM_03a,MurphyM_03b}, the main possible systematic error for the
low-$z$ results is that relative isotopic abundances may differ between the
absorption clouds and terrestrial environment/laboratory, as summarized in
Fig.~\ref{09f6}. The isotopic structures\index{Isotopic~structure} used for
Mg transitions are discussed in \cite{MurphyM_03a} whereas those for Si are
from calculations similar to \cite{BerengutJ_03a}. To our knowledge,
isotopic structures\index{Isotopic~structure} for the transitions of Fe{\sc
\,ii} are not known. However, these should be far more `compact' than those
of, say, Mg since the normal mass shift will be $>\!5$ times smaller. The
results in Figs.~\ref{09f4} \& \ref{09f5} were obtained by
fitting\index{Quasar~absorption~lines!profile~fitting} the quasar spectra
with terrestrial isotopic abundance ratios. For those systems where Mg
lines and no Si lines are fitted, these results correspond to the large
square on Fig.~\ref{09f6}, {\it vice versa} for the large triangle. The
Mg{\sc \,ii} and Si{\sc \,ii} systems approximately correspond to the low-
and high-$z$ samples respectively. If we change the heavy isotope
abundances we note the marked change in $\uDelta\alpha/\alpha$ for the Mg
systems and the comparative insensitivity for the Si systems. This is
expected given the distribution of $q$
coefficients\index{Many-multiplet~method!qcoefficients@$q$~coefficients} in
Fig.~\ref{09f2} and the diversity of transitions fitted at high-$z$ (see
\cite{MurphyM_03a} for further discussion).

In previous works we argued that the heavy isotope fraction for Mg in our
absorbers is likely to be significantly less than the terrestrial
value. This is based on observations of low
metallicity\index{Metallicity}\index{Damped~Lyman-\alpha~systems}\index{Element~abundances}\footnote{Metallicity
is the relative metal abundance with respect to that in the Solar System
environment. For LSSs and DLAs, $\log_{10} Z$ typically ranges from $-2.5$
to $-0.5$.}, $Z$, stellar environments in our Galaxy
\cite{GayP_00a,YongD_03b} and theoretical models of Galactic chemical
evolution\index{Chemical~evolution} \cite{TimmesF_95a,FennerY_03a} where
significantly sub-solar heavy isotope fractions are observed/expected at
the low $Z$s of our absorption clouds. However, some stars in
\cite{GayP_00a,YongD_03b} and those in some globular clusters
\cite{ShertoneM_86a,YongD_03a} are found to have super-solar values.

At low $Z$, the heavy Mg isotopes are thought to be primarily produced by
intermediate mass (IM; $\sim\!2\!-\!8{\rm \,M}_{\odot}$) stars in their
asymptotic giant branch\index{Asymptotic~giant~branch~stars} (AGB) phase
\cite{ForestiniM_97a,SiessL_02a,KarakasA_03a}. Reference \cite{FennerY_03a}
included a contribution from IM AGB
stars\index{Asymptotic~giant~branch~stars} in their chemical
evolution\index{Chemical~evolution} model, finding sub-solar heavy Mg
isotope abundances at low $Z$, consistent with the above
observations. Recently, \cite{AshenfelterT_03a} noted that enhanced heavy
Mg isotope abundances might be produced at low $Z$ if one assumes an
IM-enhanced stellar initial mass function\index{Initial~mass~function}
(IMF) at high $z$. However, such an IMF seems incompatible with current
constraints\index{Initial~mass~function}\index{Damped~Lyman-\alpha~systems}\index{Element~abundances}.
See \cite{KennicuttR_98a} for general discussion of the observational
constraints on the IMF. More recently, \cite{MolaroP_01a} find Ar, S and O
abundances consistent with a normal IMF in a high-$z$ DLA. See also
\cite{PettiniM_00a}. For example, an IM-enhanced
IMF\index{Initial~mass~function} could produce vast amounts of Fe via type
Ia supernovae\index{Supernovae} \cite{NomotoK_97a}, in disagreement with
Galactic and DLA\index{Damped~Lyman-\alpha~systems} abundance
studies\index{Element~abundances}. Moreover,
AGB\index{Asymptotic~giant~branch~stars} enrichment levels are constrained
by the relative element abundances\index{Element~abundances} in the
absorption clouds. For example, AGB
stars\index{Asymptotic~giant~branch~stars} produce large amounts of N
relative to other enrichment processes (e.g.~supernovae\index{Supernovae}
types Ia \& II; e.g.~\cite{DrayL_03a}). However, the abundance of N
relative to H\index{Element~abundances} is very low in
DLAs\index{Damped~Lyman-\alpha~systems}, i.e.~$10^{-3.8}$--$10^{-1.5}$
solar \cite{PettiniM_02a,CenturionM_03a}. Overall, these points are a
barrier to ad-hoc IMF\index{Initial~mass~function} changes such as those
suggested by \cite{AshenfelterT_03a}. Though enhanced heavy Mg isotope
fractions are a possible explanation of the low-$z$ varying-$\alpha$
results, we again conclude that they are an unlikely one.

\index{Isotopic~abundances|)}

\section{Varying $\alpha$ and $m_{\rm q}/\!\Lambda_{\rm QCD}$ from Atomic
  Clocks}\label{09s:clock}
\index{Variation~of~fundamental~constants!mqLambdaQCD@$m_{\rm q}/\!\Lambda_{\rm QCD}$|(}

\subsection{Introduction}\label{09ss:Cintro}
\index{Variation~of~fundamental~constants!alpha@$\alpha$|(}

The hypothetical unification\index{Grand~unification} of all interactions
implies that variation of the electromagnetic interaction constant $\alpha$
should be accompanied by variation of masses and the strong interaction
constant. Specific predictions require a model. For example, the grand
unification\index{Grand~unification} model discussed in
\cite{LangackerP_02a} (see also \cite{MarcianoW_84a,CalmetX_02a}) predicts
that the quantum chromodynamic (QCD) scale, $\Lambda_{\rm QCD}$, is
modified as
\begin{equation}\label{09QCD}
{\udelta \Lambda_{\rm QCD} \over  \Lambda_{\rm QCD}}\approx 34 {\udelta \alpha
\over \alpha}\,.
\end{equation}
The variation of quark and electron masses in this model is given by
\begin{equation}\label{09mq}
{\udelta m \over m}\sim 70 {\udelta \alpha \over \alpha}\,,
\end{equation}
resulting in an estimate for the variation of the dimensionless ratio
\begin{equation}\label{09mQCD}
{\udelta(m/ \Lambda_{\rm QCD}) \over(m/\!\Lambda_{\rm QCD})}\sim 35 {\udelta \alpha
\over \alpha}\,.
\end{equation}
The large coefficients in these expressions are generic for grand
unification\index{Grand~unification} models, in which modifications come
from high energy scales: they appear because the running strong coupling
constant and Higgs constants (related to mass) run faster than
$\alpha$. This means that if these models are correct the variation of
masses and strong interaction may be easier to detect than the variation of
$\alpha$.
 
\index{Variation~of~fundamental~constants!alpha@$\alpha$|)}

For the strong interaction there is generally no direct relation between
the coupling constants and observable quantities, unlike the case for the
electroweak forces. Since one can only measure variations in dimensionless
quantities, we must extract from the measurements constraints on variation
of $m_{\rm q}/\!\Lambda_{\rm QCD}$, a dimensionless ratio, where $m_{\rm q}$
is the quark mass (with dependence on the normalization point removed). A
number of limits on variation of $m_{\rm q}/\!\Lambda_{\rm QCD}$ have been
obtained recently from consideration of Big Bang
Nucleosynthesis\index{Big~Bang~nucleosynthesis}, quasar absorption spectra
and the Oklo natural nuclear reactor\index{Oklo~natural~fission~reactor}
which was active about 1.8 billion years ago
\cite{CowieL_95a,FlambaumV_02a,OliveK_02a,DmitrievV_03a,FlambaumV_03a,MurphyM_01d,ShlyakhterA_76a,DamourT_96a,FujiiY_00a,OberhummerH_03a,BeaneS_03a}.

Below we consider the limits which follow from laboratory atomic clock
comparisons. Laboratory limits with a time base of several years are
especially sensitive to
oscillatory\index{Variation~of~fundamental~constants!Oscillatory} variation
of fundamental constants. A number of relevant measurements have been
performed already and many more have been started or planned. The increase
in precision is very fast.

\subsection{Nuclear Magnetic Moments, $\alpha$ and
  $m_{\rm q}/\!\Lambda_{\rm QCD}$}\label{09ss:Cmu}
\index{Variation~of~fundamental~constants!alpha@$\alpha$|(}
\index{Variation~of~fundamental~constants!mu@$\mu$|(}
\index{Nuclear~magnetic~moments|(}

As pointed out by \cite{KarshenboimS_00a}, measurements of the ratio of
hyperfine structure\index{Hyperfine~structure} intervals in different atoms
are sensitive to variation of nuclear magnetic moments. The first rough
estimates of the dependence of nuclear magnetic moments on $m_{\rm
q}/\!\Lambda_{\rm QCD}$ and limits on time variation of this ratio were
obtained in \cite{FlambaumV_02a}. Using H, Cs and Hg$^+$ measurements
\cite{PrestageJ_95a}, \cite{FlambaumV_02a} limited the variation of $m_{\rm
q}/\!\Lambda_{\rm QCD}$ to about $5 \times 10^{-13}{\rm \,yr}^{-1}$. Below
we calculate the dependence of nuclear magnetic moments on $m_{\rm
q}/\!\Lambda_{\rm QCD}$ and obtain the limits from recent atomic clock
experiments with hyperfine\index{Hyperfine~structure} transitions in H, Rb,
Cs, Hg$^+$ and optical transitions in Hg$^+$. It is convenient to assume
that the strong interaction scale $\Lambda_{\rm QCD}$ does not vary, so we
will speak about variation of masses.
   
The hyperfine structure\index{Hyperfine~structure} constant can be
presented in the following form,
\begin{equation}\label{09A}
A={\rm const.} \times \left[\frac{m_{\rm e} e^4}{\hbar ^2}\right] \left[ \alpha ^2
F_{\rm rel}( Z \alpha)\right] \left[\mu \frac{m_{\rm e}}{M_{\rm p}}\right]\,.
\end{equation}
The factor in the first bracket is an atomic unit of energy. The second,
`electromagnetic', bracket determines the dependence on $\alpha$. An
approximate expression for the relativistic correction factor (Casimir
factor)\index{Casimir~factor} for the $s$-wave electron is
\begin{equation}\label{09F}
F_{\rm rel}= \frac{3}{\gamma (4 \gamma^2 -1)}\,,
\end{equation}
where $\gamma=\sqrt{1-(Z \alpha)^2}$ and $Z$ is the nuclear charge.
Variation of $\alpha$ leads to the following variation of $F_{\rm rel}$
\cite{PrestageJ_95a}:
\begin{equation}\label{09dF}
\frac{\udelta F_{\rm rel}}{F_{\rm rel}}=K \frac{\udelta \alpha}{\alpha}\,,
\end{equation}
\begin{equation}\label{09K}
K=\frac{(Z \alpha)^2 (12 \gamma^2 -1)}{\gamma^2 (4 \gamma^2 -1)}\,.
\end{equation}
More accurate numerical many-body calculations \cite{DzubaV_99b} of the
dependence of the hyperfine structure\index{Hyperfine~structure} on
$\alpha$ have shown that the coefficient $K$ is slightly larger than that
given by this formula. For Cs ($Z$=55) $K$=0.83 (instead of 0.74), for Rb
$K$=0.34 (instead of 0.29) and for Hg$^+$ $K$=2.28 (instead of 2.18).

The last bracket in (\ref{09A}) contains the dimensionless nuclear magnetic
moment, $\mu$, in nuclear magnetons (the nuclear magnetic moment
$M=\mu\times e\hbar/2 M_{\rm p} c$) and the electron--proton mass ratio,
$m_{\rm e}/M_{\rm p}$. We may also include a small correction due to the
finite nuclear size. However, its contribution is insignificant.

\index{Variation~of~fundamental~constants!gn@$g_n$|(}
\index{Variation~of~fundamental~constants!gn@$g_p$|(} Recent experiments
measured time dependence of the ratios of hyperfine
structure\index{Hyperfine~structure} intervals of $^{199}$Hg$^+$ and H
\cite{PrestageJ_95a}, $^{133}$Cs and $^{87}$Rb \cite{MarionH_03a}, and the
ratio of the optical frequency in Hg$^+$ and the $^{133}$Cs hyperfine
frequency\index{Hyperfine~structure} \cite{BizeS_03a}. In the ratio of two
hyperfine structure\index{Hyperfine~structure} constants for different
atoms, time dependence may appear from the ratio of the factors $F_{\rm
rel}$ (depending on $\alpha$) and the ratio of nuclear magnetic moments
(depending on $m_{\rm q}/\!\Lambda_{\rm QCD}$). Magnetic moments in a
single-particle approximation (one unpaired nucleon) are:
\begin{equation}\label{09mu+}
\mu=\left[g_s + (2j-1) g_l\right]/2
\end{equation}
for $j=l+1/2$ and
\begin{equation}\label{09mu-}
\mu=\frac{j}{2(j+1)}\left[-g_s + (2 j+3) g_l\right]
\end{equation}
for $j=l-1/2$. Here, the orbital $g$-factors are $g_l=1$ for valence
protons and $g_l=0$ for valence neutrons. Present values of the spin
$g$-factors, $g_s$, are $g_{\rm p}=5.586$ and $g_{\rm n}=-3.826$ for the
proton and neutron. These depend on $m_{\rm q}/\!\Lambda_{\rm QCD}$. The
light quark masses are only about 1\% of the nucleon mass [$m_{\rm
q}=(m_{\rm u}+m_{\rm d})/2 \approx 5{\rm \,MeV}$]. The nucleon magnetic
moment remains finite in the chiral limit of $m_{\rm u}=m_{\rm
d}=0$. Therefore, one may think that the corrections to $g_s$ due to the
finite quark masses are very
small. However\index{Variation~of~fundamental~constants!mpi@$m_\pi$|(},
there is a mechanism which enhances the quark mass contribution:
$\pi$-meson loop corrections to the nucleon magnetic moments which are
proportional to the $\pi$-meson mass $m_{\pi} \sim \sqrt{m_{\rm
q}\Lambda_{\rm QCD}}$. $m_{\pi}$=140 MeV is not so small.

According to \cite{LeinweberD_99a}, the dependence of the nucleon
$g$-factors on the $\pi$-meson mass can be approximated as
\begin{equation}\label{09thomas}
g(m_\pi)=\frac{g(0)}{1+ a m_\pi + b m_\pi ^2}\,,
\end{equation}
where $a=1.37{\rm \,GeV}^{-1}$, $b=0.452{\rm \,GeV}^{-2}$ for the proton
and $a=1.85{\rm \,GeV}^{-1}$, $b=0.271{\rm \,GeV}^{-2}$ for the
neutron. This leads to the following estimate:
\begin{equation}\label{09gp}
\frac{\udelta g_{\rm p}}{g_{\rm p}} =
 -0.174 \frac{\udelta m_\pi}{m_\pi}= -0.087 \frac{\udelta m_{\rm q}}{m_{\rm q}}\,,
\end{equation}
\begin{equation}\label{09gn}
\frac{\udelta g_{\rm n}}{g_{\rm n}} =
 -0.213 \frac{\udelta m_\pi}{m_\pi}= -0.107 \frac{\udelta m_{\rm q}}{m_{\rm q}}\,.
\end{equation}  
Equations (\ref{09mu+},\ref{09mu-},\ref{09gp},\ref{09gn}) give variation
of nuclear magnetic moments. For the hydrogen nucleus (proton)\,,
\begin{equation}\label{09H}
\frac{\udelta \mu}{\mu} =\frac{\udelta g_{\rm p}}{g_{\rm p}}= -0.087 \frac{\udelta m_{\rm q}}{m_{\rm q}}.
\end{equation}
For $^{199}$Hg we have the valence neutron (no orbital contribution), and
so
\begin{equation}\label{09Hg}
\frac{\udelta \mu}{\mu} =
\frac{\udelta g_{\rm n}}{g_{\rm n}} = -0.107 \frac{\udelta m_{\rm q}}{m_{\rm q}}\,.
\end{equation} 
For $^{133}$Cs we have the valence proton with $j$=7/2 and $l$=4, giving
\begin{equation}\label{09Cs}
\frac{\udelta \mu}{\mu} =
 0.22 \frac{\udelta m_\pi}{m_\pi}= 0.11 \frac{\udelta m_{\rm q}}{m_{\rm q}}\,.
\end{equation}
For $^{87}$Rb we have the valence proton with $j$=3/2 and $l$=1, resulting
in
\begin{equation}\label{09Rb}
\frac{\udelta \mu}{\mu} =
 -0.128 \frac{\udelta m_\pi}{m_\pi}=- 0.064 \frac{\udelta m_{\rm q}}{m_{\rm q}}\,.
\end{equation}
Deviation of the single-particle nuclear magnetic moment values from the
measured values is about 30\% and so we have attempted to refine them. If
we neglect the spin-orbit interaction, the total spin of nucleons is
conserved. The magnetic moment of the nucleus changes due to the spin-spin
interaction because the valence proton transfers part of its spin,
$\left<s_z\right>$, to core neutrons (transfer of spin from the valence
proton to core protons does not change the magnetic moment). In this
approximation, $g_s=(1-b)g_{\rm p} + b g_{\rm n}$ for the valence proton
(or $g_s=(1-b)g_{\rm n} + b g_{\rm p}$ for the valence neutron). We can use
the coefficient $b$ as a fitting parameter to reproduce nuclear magnetic
moments exactly. The signs of $g_{\rm p}$ and $g_{\rm n}$ are opposite,
therefore a small mixing, $b \sim 0.1$, is enough to eliminate the
deviation of the theoretical value from the experimental one. Note also
that it follows from (\ref{09gp},\ref{09gn}) that $\udelta g_{\rm p}/g_{\rm
p} \approx \udelta g_{\rm n}/{g_{\rm n}}$. This produces an additional
suppression of the mixing's effect, indicating that the actual accuracy of
the single-particle approximation for the effect of the spin $g$-factor
variation may be as good as 10\%. Note, however, that we neglected
variation of the mixing parameter $b$. This is difficult to estimate.

\index{Variation~of~fundamental~constants!gn@$g_n$|)}
\index{Variation~of~fundamental~constants!gn@$g_p$|)}
\index{Variation~of~fundamental~constants!mu@$\mu$|)}
\index{Variation~of~fundamental~constants!mpi@$m_\pi$|)}
\index{Nuclear~magnetic~moments|)}

\subsection{Results}\label{09ss:Cresu}

We can now estimate the sensitivity of the ratio of the
hyperfine\index{Hyperfine~structure} transition frequencies to variations
in $m_{\rm q}/\!\Lambda_{\rm QCD}$. For $^{199}$Hg and hydrogen we have
\begin{equation}\label{09HgH}
\frac{\udelta[A({\rm Hg})/A({\rm H})]}{[A({\rm Hg})/A({\rm H})]}= 2.3
\frac{\udelta \alpha}{\alpha} -0.02 \frac{ \udelta [m_{\rm q}/\!\Lambda_{\rm
      QCD}]}{[m_{\rm q}/\!\Lambda_{\rm QCD}]}\,.
\end{equation}
Therefore, the measurement of the ratio of Hg and hydrogen
hyperfine\index{Hyperfine~structure} frequencies is practically insensitive
to the variation of light quark masses and the strong
interaction. Measurements \cite{PrestageJ_95a} constrain variations in the
parameter $\tilde{\alpha}= \alpha [m_{\rm q}/\!\Lambda_{\rm QCD}]^{-0.01}$:
\begin{equation}\label{09limitHgH}
\left|\frac{1}{ \tilde{\alpha}}\frac{\D\tilde{\alpha}}{\D t}\right|
 < 3.6 \times 10^{-14}{\rm \,yr}^{-1}\,.
\end{equation}
Other ratios of hyperfine\index{Hyperfine~structure} frequencies are more
sensitive to $m_{\rm q}/\!\Lambda_{\rm QCD}$. For $^{133}$Cs and $^{87}$Rb we
have
\begin{equation}\label{09CsRb}
\frac{\udelta[A({\rm Cs})/A({\rm Rb})]}{[A({\rm Cs})/A({\rm Rb})]}= 0.49
 \frac{\udelta \alpha}{\alpha} +0.17 \frac{ \udelta [m_{\rm q}/\!\Lambda_{\rm
 QCD}]}{[m_{\rm q}/\!\Lambda_{\rm QCD}]}\,.
\end{equation}
\begin{sloppypar}
Therefore, measurements \cite{MarionH_03a} constrain variations in the
parameter $X=\alpha^{0.49} [m_{\rm q}/\!\Lambda_{\rm QCD}]^{0.17}$:
\end{sloppypar}
\begin{equation}\label{09limitCsRb}
 \frac{1}{X}\frac{\D X}{\D t}=(0.2 \pm 7) \times 10^{-16}{\rm \,yr}^{-1}\,.
\end{equation}
Note that if the relation (\ref{09mQCD}) is correct, variation of $X$ would
be dominated by variation of $m_{\rm q}/\!\Lambda_{\rm QCD}$: (\ref{09mQCD})
would give $X \propto \alpha^{7}$.

For $^{133}$Cs and H we have
\begin{equation}\label{09CsH}
\frac{\udelta[A({\rm Cs})/A({\rm H})]}{[A({\rm Cs})/A({\rm H})]}=0.83
 \frac{\udelta \alpha}{\alpha} +0.2 \frac{ \udelta [m_{\rm
 q}/\!\Lambda_{\rm QCD}]}{[m_{\rm q}/\!\Lambda_{\rm QCD}]}\,.
\end{equation}
Therefore, measurements \cite{DemidovN_93a,BreakironL_93a} constrain
variations of the parameter $X_H= \alpha^{0.83} [m_{\rm q}/\!\Lambda_{\rm
QCD}]^{0.2}$:
\begin{equation}\label{09limitCsH}
\left|\frac{1}{X_H}\frac{\D X_H}{\D t}\right|< 5.5 \times 10^{-14}{\rm
\,yr}^{-1}\,.
\end{equation}
If we assume the relation (\ref{09mQCD}), we would have $X_H \propto
\alpha^{8}$.

The optical clock transition energy, $E({\rm Hg})$, at $\lambda=282{\rm
\,nm}$ in the Hg$^+$ ion, can be presented as
\begin{equation}\label{09E}
E({\rm Hg})={\rm const.} \times \left[\frac{m_{\rm e} e^4}{\hbar ^2}\right]
F_{\rm rel}(Z \alpha)\,.
\end{equation}
Note that the atomic unit of energy (first bracket) is cancelled out and so
we do not consider its variation. Numerical calculation of the relative
variation of $E({\rm Hg})$ has given \cite{DzubaV_99b}
\begin{equation}\label{09dE}
\frac{\udelta E({\rm Hg})}{E({\rm Hg})}=-3.2\frac{\udelta \alpha}{\alpha}\,.
\end{equation}
Variation of the ratio of the Cs hyperfine\index{Hyperfine~structure}
splitting $A({\rm Cs})$ to this optical transition energy is equal to
\begin{equation}\label{09CsHgE}
\frac{\udelta[A({\rm Cs})/E({\rm Hg})]}{[A({\rm Cs})/E({\rm Hg})]}= 6.0
\frac{\udelta \alpha}{\alpha} + \frac{ \udelta [m_{\rm e}/\!\Lambda_{\rm
QCD}]}{[m_{\rm e}/\!\Lambda_{\rm QCD}]} +0.11 \frac{ \udelta [m_{\rm
q}/\!\Lambda_{\rm QCD}]}{[m_{\rm q}/\!\Lambda_{\rm QCD}]}\,.
\end{equation}
Here we have taken into account that the proton mass $M_{\rm p} \propto
\Lambda_{\rm QCD}$. The factor 6.0 before $\udelta \alpha$ appeared from
$\alpha^2 F_{\rm rel}$ in the Cs hyperfine\index{Hyperfine~structure}
constant (i.e.~2+0.83) and $\alpha$-dependence of $E({\rm Hg})$
(i.e.~3.2). Therefore, the results of \cite{BizeS_03a} give the limit on
variation of the parameter $U= \alpha^{6} [m_{\rm e}/\!\Lambda_{\rm
QCD}][m_{\rm q}/\!\Lambda_{\rm QCD}]^{0.1}$:
\begin{equation}\label{limitCsHgE}
\left|\frac{1}{U}\frac{\D U}{\D t}\right|< 7 \times 10^{-15}{\rm \,yr}^{-1}\,.
\end{equation}
If we assume the relation (\ref{09mQCD}), we would have $U \propto
\alpha^{45}$.
Note that we present such limits on
$\left|(\D \alpha/\D t)/\alpha\right|$ as illustrations only since they are
strongly model-dependent.

\index{Variation~of~fundamental~constants!alpha@$\alpha$|)}
\index{Variation~of~fundamental~constants!mqLambdaQCD@$m_{\rm q}/\!\Lambda_{\rm QCD}$|)}

\section{Conclusions}\label{09s:conc}
\index{Variation~of~fundamental~constants!alpha@$\alpha$|(}

We have presented evidence for a varying $\alpha$ based on many-multiplet
measurements in 143 Keck/HIRES quasar absorption systems covering the
redshift range $0.2 < z_{\rm abs} < 4.2$: $\uDelta\alpha/\alpha = (-0.57
\pm 0.11) \times 10^{-5}$. Three independent observational samples give
consistent results. Moreover, the low- and high-$z$ samples are also
consistent, which cannot be explained by simple systematic errors
(Fig.~\ref{09f2}). Our results therefore seem internally robust. The
possibility that the isotopic abundances\index{Isotopic~abundances} are
very different in the absorption clouds and the laboratory is a potentially
important systematic effect. A high heavy isotope fraction for Mg
($\Gamma^{25,26}_{\rm Mg}\approx 0.5$) compared with the terrestrial value
($\Gamma^{25,26}_{\rm Mg}\approx 0.21$) may explain the low-$z$ results
(Fig.~\ref{09f6}). However, observations of
low-metallicity\index{Metallicity} stars and Galactic chemical
evolution\index{Chemical~evolution} (GCE) models suggest sub-solar values
of $\Gamma^{25,26}_{\rm Mg}$ in the quasar absorption systems. GCE models
with a stellar initial mass function\index{Initial~mass~function} greatly
enhanced at intermediate masses may produce large quantities of heavy Mg
isotopes via asymptotic giant branch stars. However, such models disagree
with the observed element abundances\index{Element~abundances} in quasar
absorption systems. The high-$z$ results are insensitive to the isotopic
fraction of $^{29,30}$Si. However, we stress the need for theoretical
calculations and laboratory measurements of isotopic
structures\index{Isotopic~structure} for other elements/transitions
observed in quasar absorption systems.

Aside from a varying $\alpha$, no explanation of our results currently
exists which is consistent with the available observational evidence. The
results can best be refuted with detailed many-multiplet analyses of quasar
absorption spectra from different telescopes/instruments now available
(e.g.~VLT/UVES, Subaru/HDS).

\index{Variation~of~fundamental~constants!mqLambdaQCD@$m_{\rm q}/\!\Lambda_{\rm QCD}$|(}

We have calculated the dependence of nuclear magnetic
moments\index{Nuclear~magnetic~moments} on quark masses. This leads to
limits on possible variations in $m_{\rm q}/\!\Lambda_{\rm QCD}$ from
recent laboratory atomic clock experiments involving
hyperfine\index{Hyperfine~structure} transitions in H, Rb, Cs, Hg$^{+}$ and
an optical transition in Hg$^{+}$. These limits can be compared with limits
on $\alpha$-variation within the context of grand
unification\index{Grand~unification} theories. Unfortunately, this
comparison is strongly model-dependent. See, for example, \cite{DentT_03b}.

\index{Variation~of~fundamental~constants!alpha@$\alpha$|)}
\index{Variation~of~fundamental~constants!mqLambdaQCD@$m_{\rm q}/\!\Lambda_{\rm QCD}$|)}


\label{09_}

\end{document}